\documentclass[12pt]{article}  
\usepackage{natbib, emulateapj5, graphics}

\slugcomment{Submitted to Astrophysical Journal}
\shortauthors{Gawiser et al.}
\shorttitle{Damped Ly $\alpha$ Absorbers at $z\simeq 4$}
\begin{document}

\title{First Investigation of the Clustering Environment of Damped 
Lyman $\alpha$ Absorbers at $z\simeq 4$$^1$}




\author{Eric Gawiser\altaffilmark{2,3}, 
Arthur M. Wolfe\altaffilmark{2,3},
Jason X. Prochaska\altaffilmark{2,4}, 
Kenneth M. Lanzetta\altaffilmark{2,5}, 
Noriaki Yahata\altaffilmark{5}, 
\& Andreas Quirrenbach\altaffilmark{2,3}}

\altaffiltext{1}{Based 
on data obtained at the W.~M.~Keck Observatory which is operated as a 
scientific partnership among the University of California, the 
California Institute of Technology, and NASA and was made
possible by the generous financial support of the W.~M.~Keck Foundation.}
\altaffiltext{2}{Visting Astronomer, Keck Observatory}
\altaffiltext{3}{Center 
for Astrophysics and Space Sciences and 
Department of Physics, University of California at San Diego,
 La Jolla, CA  92037, egawiser,awolfe,aquirrenbach@ucsd.edu}
\altaffiltext{4}{The Observatories of the Carnegie Institute 
of Washington, 813 Santa Barbara Street, Pasadena, CA  91101, 
xavier@ociw.edu}
\altaffiltext{5}{Department of Physics and Astronomy, 
State University of New York at Stony Brook, Stony Brook, NY  11794-3800,
lanzetta,nyahata@sbastr.ess.sunysb.edu}


\begin{abstract}
We report the first observations of the clustering environment of 
damped Lyman $\alpha$ absorption systems at $z\simeq 4$.  
Color selection and photometric redshifts were used to select 
44 candidate Lyman-break galaxies brighter than I$_{AB}=25.5$ 
from deep BRI images 
of the 35 arcmin$^2$ field containing the quasar BR 0951$-$04.   
Multislit spectroscopy of 35 candidate galaxies was performed 
and 8 of these candidates have been confirmed as $z>3.5$ 
Lyman-break galaxies.   
With only BRI photometry, the photometric redshifts are quite 
accurate for the spectroscopically confirmed galaxies but have a 
high rate of misclassification due to color degeneracies between
Lyman-break galaxies and low-redshift ellipticals.    
Both of the $z>3.5$ galaxies found within 
$15''$ of the quasar line-of-sight appear to be causing 
absorption systems in the quasar spectrum.  
We use a battery of statistical tests to look for clustering in the 
redshift histogram of the $z>3.5$ galaxies but do not find 
measurable clustering of these Lyman-break galaxies with the 
damped Lyman $\alpha$ absorbers.
With a larger sample of galaxies, our 
method should determine 
the cross-correlation between 
these objects, which probes the bias and hence the mass of the 
damped Lyman $\alpha$ absorbers.   
\end{abstract}


\maketitle

\section{Introduction}
\label{sect:intro}

Galaxy formation is one of the great unsolved problems in 
astrophysics.  We have clear evidence that many galaxies were 
in place at redshifts of $3-4$ but little knowledge of
how or when they formed. 
Using Lyman-break imaging techniques, 
\citet{steideletal96}
discovered a large sample of  $z$ $\simeq$ 3 galaxies with 
rest-frame ultraviolet luminosities   
$L_{UV}$ $>$ 10$L_{*}$  and 
comoving number density comparable to that of $L_*$ galaxies today.
These Lyman-break galaxies (LBGs) 
exhibit surprisingly strong clustering for 
$z \simeq 3$, with overdensities $\delta \rho/ \rho$ $>$ 3 
and comoving correlation length $r_{0}$ $\approx$ 5$h^{-1}$ Mpc 
\citep{adelbergeretal98}.  In hierarchical models of  
structure formation such as $\Lambda$CDM, 
the LBGs will undergo significant 
merging and should be even more clustered, rare massive galaxies today.

The damped Lyman $\alpha$ absorption systems (DLAs), 
on the other hand, are thought to be the progenitors of typical 
present-day galaxies 
\citep{wolfeetal86,kauffmann96}.  
We therefore 
must learn about the mass and environment of high-redshift 
DLAs if we want to understand the formation of typical galaxies like the 
Milky Way.  
The damped Lyman $\alpha$ absorbers are dense clouds of neutral hydrogen 
seen in absorption along the line of sight to distant quasars.
They appear to be drawn from the bulk of the
protogalactic mass distribution;  this conclusion is supported by 
the agreement between the comoving mass density  of neutral gas
at $z>2$ and the mass density of visible stars in current galaxy disks 
\citep{wolfeetal95}.
However, the total gas content cannot distinguish CDM scenarios 
in which DLAs are numerous low-mass protogalaxies from 
passive evolution models in which the DLAs are 
more rarely occurring massive objects.
Determining the mass of individual DLAs is also critical for testing 
models of galaxy formation within the CDM paradigm.  
This determination cannot be 
performed using the 
absorption characteristics of the DLAs, as the comoving density 
of DLAs cannot be determined without knowing their typical 
cross-sectional area.  
One way to distinguish among models for the DLAs is 
to measure the kinematics of 
the gas \citep{ prochaskaw97, prochaskaw98,wolfep00a, wolfep00b}.  
This paper describes an independent
method which seeks to determine the mass of the DLAs by investigating 
the extent to which they cluster with LBGs.  
 Since DLAs are predicted by hierarchical cosmologies 
to be rather weak in emission, and observation is consistent 
with this \citep{steidelph95}, 
it may be easier to constrain their mass using 
large-scale structure techniques than by searching for 
their emission (see \citealt{wolfe93}).  

Section \ref{sect:lss} describes the theoretical basis for our investigation 
of the cross-correlation of DLAs and LBGs.  Imaging 
observations and photometric data reduction 
are described in \S \ref{sect:imaging},  
and our multislit spectroscopy is presented in \S \ref{sect:spectroscopy}.
We discuss the search for emission from the DLAs themselves in 
 \S \ref{sect:emission}, search for evidence of clustering with nearby 
LBGs in \S \ref{sect:clustering}, and discuss 
our results in \S \ref{sect:conc}.  

\section{Large-Scale Structure at High Redshift}
\label{sect:lss}

Hierarchical models of structure formation predict 
that the most massive galaxies tend to form in regions of unusually high 
density, whereas low-mass galaxies are more uniformly distributed in 
space.  This leads to an enhancement in the clustering of high-mass 
galaxies, referred to as bias,  
since they typically form near each other in rare overdense regions.
Clustering thus provides a way to probe the mass 
distribution of DLAs; if there is a significant overdensity  
of LBGs near 
damped Lyman $\alpha$ absorbers 
it indicates that the DLAs 
reside in massive dark halos.  If, on the 
other hand, the DLAs are preferentially 
found in regions with an underdensity 
of LBGs, it would 
argue that some environmental effect, perhaps ionizing radiation from massive 
stars in the LBGs (see \citealt{steidelpa00}), 
prevents the occurrence of large column densities of 
neutral hydrogen in overdense regions.  The expected result in
hierarchical cosmologies is intermediate; DLAs are predicted to be 
low-mass objects and should be found in both the (proto)clusters 
and the field with 
typically only a small enhancement of LBGs nearby.  

While the cross-correlation of galaxies and clusters has 
been studied at low redshift (e.g. \citealt{mopx93}), these objects can 
also be studied through their autocorrelation.  The inevitable sparse 
sampling of DLAs and other absorption systems caused by 
only being able to detect them on the line of sight towards quasars or 
bright galaxies makes it very difficult to measure their 
autocorrelation.  Hence, the cross-correlation is the best, and 
perhaps only, way to study the bias of DLAs.   
Another advantage is that the cross-correlation of DLAs and LBGs 
can be established without detecting emission from 
the galaxy causing the damped Lyman $\alpha$ absorption.  

The probability of finding an LBG in a small volume $dV$ a distance 
$r$ from a known DLA is enhanced versus the average density of LBGs, 
\begin{equation}
dP_{LBG} = [1 + \xi_{DL}(r)] \rho_{LBG} dV		\; , \;  \; \; 
\label{eq:density}
\end{equation}
where $\xi_{DL}$  is the cross-correlation function of DLAs and LBGs.  
On scales where linear bias is a good model, 
the cross-correlation of any two classes of objects is determined 
by multiplying the dark matter correlation function $\xi(r)$ by the 
product of the bias factors of each class:  
\begin{equation}
\xi_{12}(r) = b_1 b_2 \xi(r)\; . \; \; \; 
\label{eq:cross}
\end{equation}
The bias of a set of collapsed objects of a given mass can be determined 
using the extended Press-Schechter method 
(see \citealt{mojw96}
).  
The autocorrelation of LBGs ($\xi_{LL}$) is known at 
$z=3$ \citep{adelbergeretal98} but has not yet been determined 
 at $z=4$ \citep[hereafter S99]{steideletal99}, probably due to the 
increased difficulty of obtaining redshifts at $z \simeq 4$.  
The bias of other high-redshift objects such as quasars, radio galaxies, or 
Lyman $\alpha$ emitters can also be determined using the 
cross-correlation function.

\section{Imaging and Photometry}
\label{sect:imaging}

We took deep BRI images of the field around the $z=4.37$ quasar BR 0951$-$04 
\citep{storrielombardietal96} 
with the LRIS instrument \citep{okeetal95} 
at the Keck Observatory.  This field was chosen because there 
are two DLAs along the line of 
sight towards BR 0951$-$04, at $z=3.86$ and $z=4.20$ 
\citep{storrielombardiw00}.   
Table \ref{tab:obs} lists the exposure time and final depth in each filter.  
Images were reduced using a mix of standard IRAF routines and customized 
code.  The final images cover a $5'\times7'$ field.  
Photometry was performed on the R image using 
SExtractor \citep{bertina96} and then 
a custom code was used to extract photometry from the final B and I images 
using the exact apertures selected in R.  
Prochaska et al. (in preparation) 
describe our data reduction methods in greater detail and 
make our photometric catalog and images public.

We modified the $G \Re i$ color selection criteria of S99 for our 
Johnson-Cousins BRI filter set.\footnote{Our 
filter set corresponds closely 
to their custom set except for an overall 
200\AA ~($\Delta z \simeq 0.2$) shift 
bluewards in B and R, allowing the approximation (B$-$R)$_{AB} =$($G- \Re$).  
Because our I filter is close to theirs in effective wavelength, 
our (R$-$I)$_{AB}$ 
colors should be roughly 1.6 times their ($\Re -i$) colors.}
Objects located in the region of color-color space bounded by 
(B$-$R)$_{AB} \geq 2.0$, (B$-$R)$_{AB} \geq 1.2$(R$-$I)$_{AB} +1.6$, 
and (R$-$I)$_{AB} \leq 1.0$, 
have a high probability of being Lyman-break galaxies at $z > 3.5$. 
We have also used photometric redshifts 
(see \citealt{yahataetal00})
and have found at least one LBG that would have been missed using 
only the two-color region of 
S99.  Almost all of the color-selected candidates are 
identified by the photometric redshift technique as having a high probability 
of being at $3.5<z<4.5$. 
The photometric redshift method accounts for the individual 
photometric uncertainties in the B,R, and I 
fluxes of each object and compares the 
object fluxes with a set of template spectra rather than a 
single two-color region.    
Carefully designed seven-color photometric redshift schemes yield 
a scatter of $\Delta z = 0.2$ at these redshifts \citep{hoggetal98}.   
It should therefore be impossible for photometric redshift selection 
to create a redshift distribution as narrow 
as the $\Delta z \simeq 0.04$ ($\simeq$2000 km/s) redshift spikes seen by
 \citet{adelbergeretal98}.  

We visually inspected all candidate 
galaxies in our images and rejected roughly half of the initial candidates for 
having suspect photometry.  Many of the rejected objects 
lied near the edge of our imaging field in regions of lower signal-to-noise.  
It is no surprise that such a high fraction of the initial candidates 
had suspect photometry, as the color selection region is sparsely populated 
and therefore the small fraction of low-redshift galaxies scattered 
into the color-selection region by photometric errors 
can easily comprise a large fraction of the candidates.  
Bright (I$<23$) objects with clearly non-stellar morphology were rejected as 
low-redshift interlopers, but bright 
stellar objects were retained in case they 
were new quasars or lensing-magnified LBGs; 
most LBGs will look similar to point sources 
at the resolution of our images.  

\section{Spectroscopy and Redshift Determination}
\label{sect:spectroscopy}

We performed longslit spectroscopy of two objects near BR 0951$-$04 
and multislit spectroscopy of this field with four different slitmasks, 
all with 1.5$''$ width slitlets and 
the 150 line/mm grating (30 \AA ~resolution) on Keck-LRIS. 
Table \ref{tab:obs} lists details of our observations.  
Each slitmask targeted about 15 candidate objects.
The entire LRIS imaging field was covered using 2 slitmasks 
(termed left and right) that overlap in the region near the quasar.  These 
masks were redesigned for the March 2000 run to eliminate slits on objects 
whose redshifts were already determined and to 
add slits on additional candidate LBGs.
We used a GG495 dichroic filter to block second order light 
in December 1999 but decided against this 
in the subsequent run because
wavelengths below 4950 \AA ~can be used to 
check for the presence of the Lyman limit.  
The spectroscopic data were reduced using standard IRAF techniques called 
by a customized version of the BOGUS 
code\footnote{Available at http://zwolfkinder.jpl.nasa.gov/$\sim$stern/homepage/bogus.html}.  The functions {\it response} 
(halogen lamp spectrum fitting), {\it background} (sky subtraction) and 
{\it apall} 
(spectrum extraction) were performed interactively following the 
multislit data reduction cookbook 
available at http://mamacass.ucsd.edu/people/gawiser/cookbooks.html .
Redshifts were determined interactively using a customized PGPLOT routine 
which compared the measured spectra with templates of expected absorption 
(and emission) lines for LBGs \citep{steidelpa00,lowenthaletal97}
and E+A galaxies \citep{zabludoffetal96}.  
Gawiser et al. (in preparation) describe our spectroscopic data reduction 
in greater detail.  

We obtained spectra of 35 candidate galaxies in the field of BR 0951$-$04
(see Table~\ref{tab:q0951}).  
Eight candidates turn out to be galaxies at $z>3.5$, thirteen appear to be  
low-redshift interlopers (typically ellipticals 
at $z \simeq 0.5$) and fourteen have inconclusive spectra.   
Our color-selected sample with I$_{AB}\leq 25$ is directly comparable 
to the objects observed by S99.  We find 0.5 candidates per sq. arcmin, 
twice the candidate density of S99;  the difference is 
unlikely to be caused solely by Poisson fluctuations 
given that there are 19 objects in this 
category in Table~\ref{tab:q0951}.  Our rate of determining redshifts 
for objects observed spectroscopically is comparable to S99, and the 
fraction of candidates identified as interlopers, i.e. at $z<3.5$, 
is higher but very sensitive to small-number statistics at present.  
We will revisit this comparison once a larger sample of BRI-selected 
galaxies is available.  

\begin{figure}[h!]
\begin{center}
\scalebox{0.9}[0.9]{
{\includegraphics{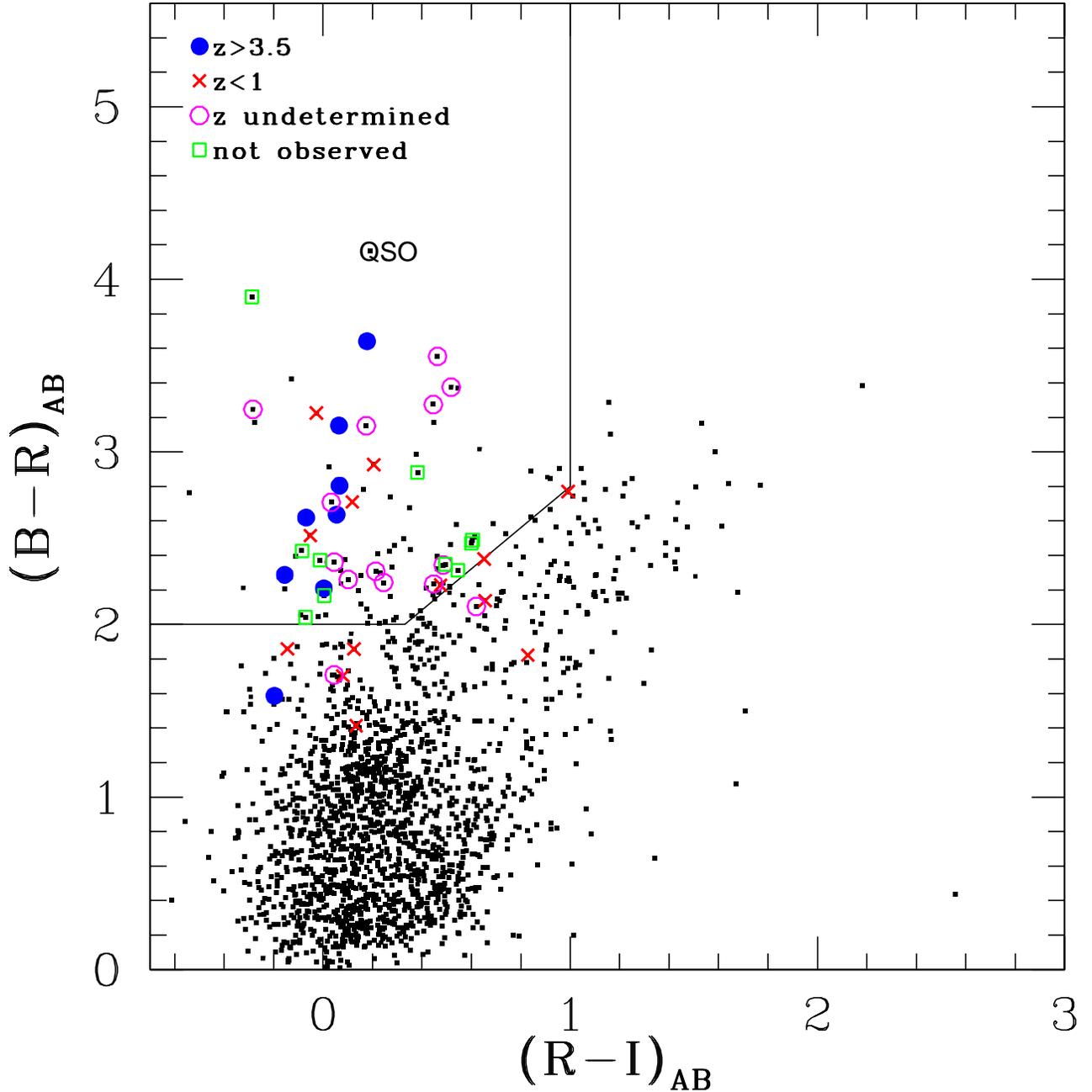}}}
\caption{\small 
Two-color diagram of objects imaged 
in field of BR 0951$-$04 with I$_{AB}\leq 25.5$ (with the addition of 
object G1).    
Objects that were selected as candidate 
LBGs based on their colors and/or photometric redshifts have been 
assigned large symbols as identified in the upper left corner.  
Objects that fall in the color selection region but are represented 
only by small points were found by visual
 inspection to have suspect photometry 
and were therefore not considered as candidate LBGs.
\label{fig:2clr}
}
\end{center}
\end{figure}

Figure~\ref{fig:2clr} shows the two-color diagram of all 2254 objects 
with I$_{AB}<25.5$ detected in the field with the status of LBG candidates 
shown.  Table~\ref{tab:gals} lists photometry and photometric redshift 
information on each $z>3.5$ galaxy with its best-fit redshift and 
uncertainty.  The photometric redshifts based on BRI alone are impressively 
accurate for the confirmed high-redshift galaxies, with a dispersion of 
$\sigma = 0.18$, including three cases where $z_{phot}$ differs from the 
spectroscopic redshift by only 0.01.  However, adding in the catastrophic 
failures where $z\simeq 0.5$ galaxies were assigned photometric 
redshifts $z_{phot} \simeq 4$ would 
drastically increase this dispersion.  Adding near-infrared (JHK) or 
optical Z-band photometry would be quite helpful in discriminating against 
these intrinsically-red low-redshift interlopers.  

The increased fraction of interlopers among candidates 
selected primarily by photometric redshift confirms that the 
S99 color selection region is the most efficient 
place to find LBGs.
However, we differ from S99 in one major 
respect.  We do not assume that the 
spectra with insufficient signal-to-noise for redshift determination
are drawn from the same population of objects as those with 
identified redshifts.  Thus there is a factor of two uncertainty 
in the number abundance of color-selected 
$z>3.5$ galaxies, both here and in S99, caused by almost half of 
the candidates lacking redshift identifications.
  Extrapolating from the observed color distribution 
of LBGs at $z=3$, S99 predict that roughly half of the $z=4$ LBGs will lie 
outside of the color selection region.  The failure of photometric redshifts 
to find these LBGs implies either that this incompleteness has been 
overestimated or that the photometric redshift technique needs to be modified 
to better identify LBGs with unusual colors.  

\begin{figure}[t!]
\begin{center}
\scalebox{0.95}[0.95]{\rotatebox{0}{\includegraphics{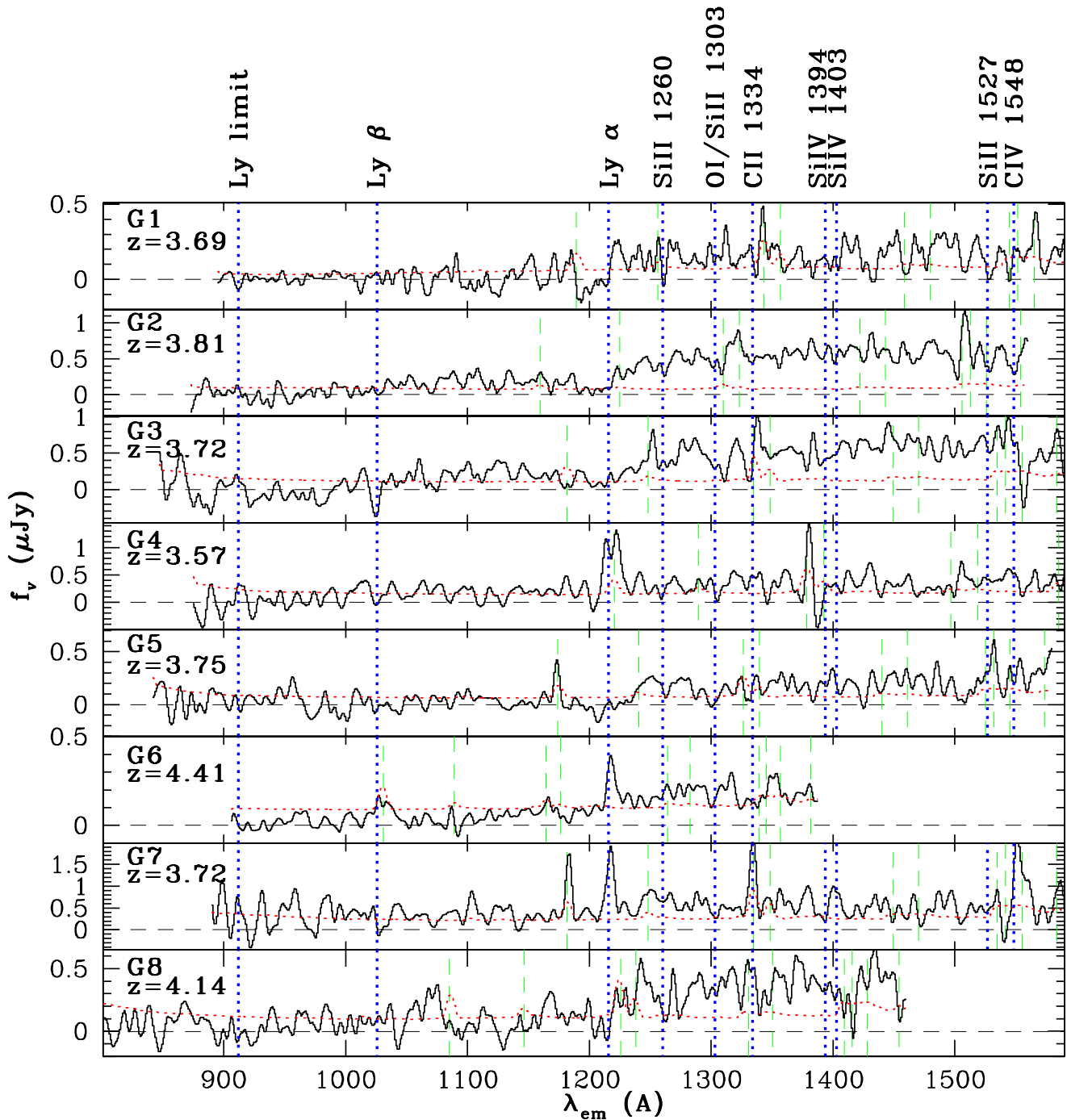}}}
\caption{\small 
Spectra of 8 confirmed $z>3.5$ galaxies in the field of BR 0951$-$04 after  
being smoothed by the resolution element.  
The dotted spectrum shows 
the $1 \sigma$ uncertainty.  
Spectra are plotted versus inferred rest-frame wavelength, 
with dotted vertical lines showing expected absorption/emission features.  
Flux is shown in units of $\mu$Jy ($1 \mu$Jy $= 10^{-29}$ erg cm$^{-1}$ 
s$^{-1}$ Hz$^{-1}$).  
Dashed vertical lines show sky lines (at 5577, 5892, 6300, 6364, 6840, 
6940, 7244, 7276, 7340, and 7475 \AA ~) that 
are too strong to be subtracted well and 
can be used as a rough guide to the observed wavelengths.  Observed 
wavelengths from 4000 to 7500 \AA ~ are shown; the signal-to-noise degrades 
considerably below that due to reduced grating throughput and 
CCD quantum efficiency and above that due 
to sky emission bands.  Object G6 was observed with a GG495 dichroic, so 
only wavelengths above 4900 \AA ~ are shown.
\label{fig:spectra}
}
\end{center}
\end{figure}

Figure~\ref{fig:spectra} 
shows the rest-frame spectra of the confirmed $z>3.5$ galaxies.   
These spectra show the typical mix of Lyman $\alpha$ emission 
(G4,G6,G7) and Lyman $\alpha$ absorption (G1,G2,G8) as well as 
two vague suggestions of emission within a broad absorption trough (G3,G5).  
The signal-to-noise is generally quite poor, illustrating why I=25 is 
usually considered the limit for ground-based optical spectroscopy.

\section{Search for Emission from Damped Lyman $\alpha$ Absorbers}
\label{sect:emission}

At $z \sim 4$, Lyman-break galaxies found
within a 10$''$ radius of the quasar may be the source of 
damped Lyman $\alpha$ absorption.  
This radius corresponds to the maximum impact parameter ($\sim 100$ kpc)
inferred from absorption line
statistics in the passive evolution model \citep{storrielombardiw00, 
bunkeretal99}.
If DLAs represent formed disk galaxies as postulated by 
\citet{wolfe97},
we should be able to detect emission from up to $20\%$ of them
at R$ \leq 25$ \citep{malleretal00}.
However, 
in hierarchical models of structure formation,
a much higher fraction of 
DLAs are expected to be too dim and/or too close to the quasar 
to yield detectable emission 
\citep{haehneltsr98,haehneltsr00}.
Our survey hopes to differentiate between these hypotheses; if emission 
from DLAs is found repeatedly 
at large impact parameters or high luminosity it will 
represent
 a serious conflict with the predictions of hierarchical cosmology.  

We performed a complete survey for possible emission from 
DLAs on the line of sight toward the 
quasar down to I$_{AB} = 26$ at angular distances up to $10''$, although 
the PSF of the quasar hampers our survey at angles less than $2''$ where   
hierarchical cosmology predicts that the emission from most DLAs will 
be found.  
We identified three objects within $10''$ of the QSO 
line-of-sight 
that had a reasonable chance of being at high redshift (our 
selection criteria were less stringent in this region).  
One of these, G1, meets the color selection 
criteria but has I$_{AB}=25.7$ 
and appears to have a redshift of $3.69$.  Although 
G1 is $10''$ from the QSO, it may be responsible for 
the $z=3.703$ absorption system seen in SiIV and CIV 
\citep{storrielombardietal96}.  
The second nearby object appears to be a low-redshift interloper, 
and the third has inconclusive redshift.
Another color-selected galaxy, G2 at $z=3.81$, lies $15''$ ($\simeq 200$ kpc) 
from the QSO and may be responsible for the $z=3.818$ absorption system seen 
in Lyman $\alpha$ and CIV in our Keck-HIRES spectrum of 
BR 0951$-$04 \citep{prochaskaw99}.  
It will be interesting to determine the fraction 
of galaxies detected in both emission and absorption as a function of 
distance from the quasar line-of-sight.
We have not detected emission physically associated with 
either of the damped Lyman $\alpha$ absorbers 
towards BR 0951$-$04 spectroscopically, but 
all of the models predict that a larger sample is needed before we should 
expect to find an LBG physically associated with a DLA.

\section{Clustering Analysis}
\label{sect:clustering}

\begin{figure}[b!]
\begin{center}
\scalebox{0.9}[0.9]{
{\includegraphics{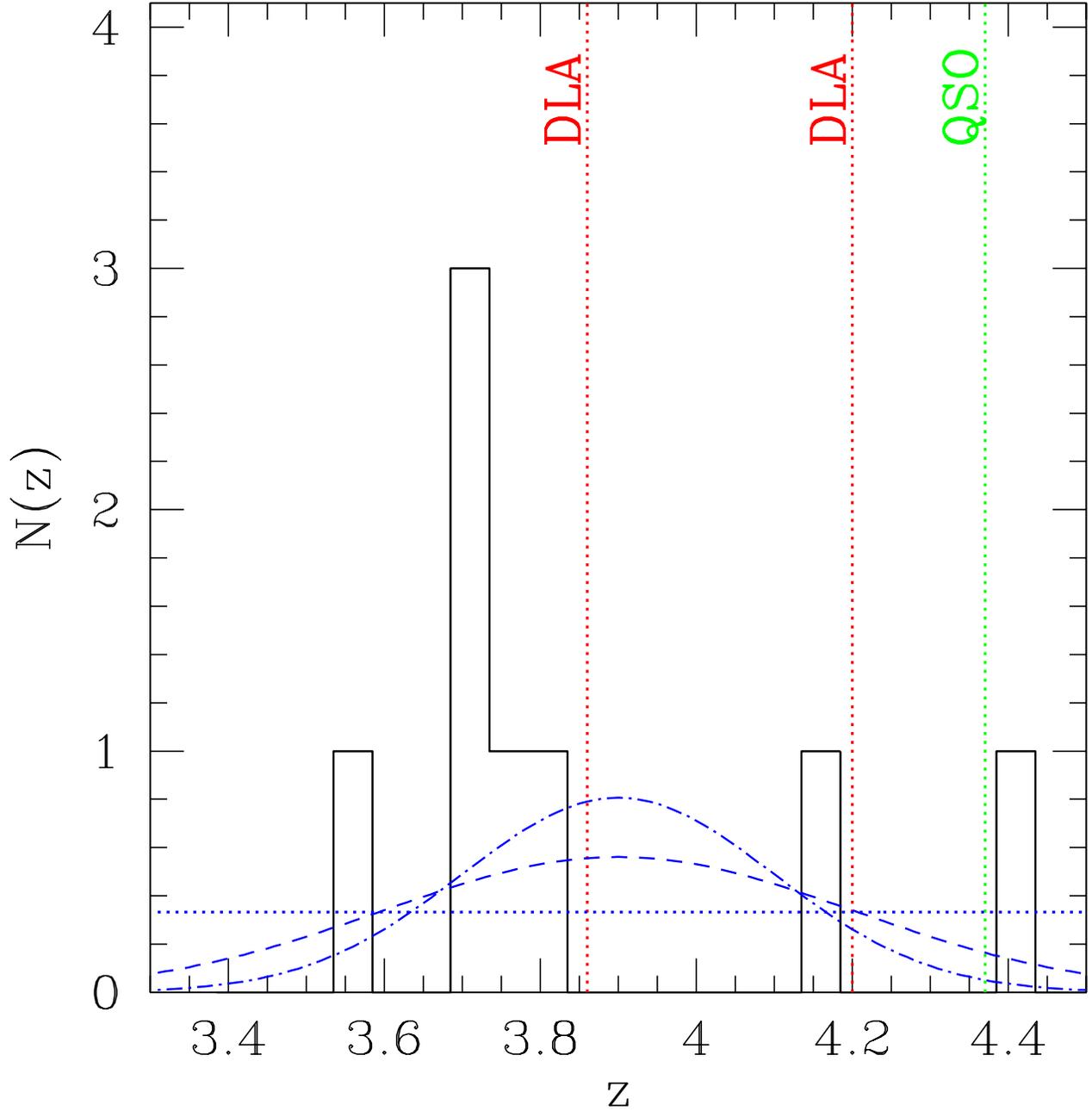}}}
\caption{\small 
Histogram of redshifts of LBGs in field of BR 0951$-$04.  
Redshifts of DLAs and quasar 
are shown by vertical dotted lines.  The range of possible 
expected redshift distributions described in the text is also plotted;
the dot-dashed line has mean $z=3.9$, $\sigma=0.2$, the dashed 
line has mean $z=3.9$, $\sigma = 0.3$, and the dotted line is 
a uniform distribution from $z=3.3$ to $z=4.5$.  All are normalized 
to produce 8 galaxies.  
\label{fig:zhist}
}
\end{center}
\end{figure}

Figure~\ref{fig:zhist} shows the redshift histogram of objects in 
the field of BR 0951$-$04, with a binsize of $\Delta z = 0.05$ 
($\sim 10 h^{-1}$ Mpc) chosen 
because most of our redshift uncertainties are smaller than 
this value.  
We chose to place the $z=3.86$ DLA 
in the center of a bin in order to maximize the chance of finding 
an overdensity (or underdensity) 
of LBGs near the DLA if one exists.  However, the results described 
below are not 
altered by shifting the bin locations.

There are eight confirmed LBGs in the BR 0951$-$04 field, 
so the search for clustering is plagued by small number statistics.  
Before trying to determine the amplitude of a cross-correlation function, 
it makes sense to ask if any clustering has been observed that is inconsistent 
with poisson fluctuations about the expected redshift distribution of 
LBGs.  The expected redshift distribution refers to the multiplication of the 
true redshift distribution of LBGs on an average line of sight times the 
observational selection function.
While the observational selection function can be simulated (see S99), the 
true redshift distribution of LBGs is unknown, so it is necessary 
to determine the expected redshift distribution empirically.  
We have not yet observed a sufficient number of LBGs to determine how our 
expected redshift distribution differs from that of S99, who found 
$<z>=4.13, \sigma=0.26$.
  We do expect a shift of 0.2 in redshift, so we choose to 
bracket the possibilities by showing three possible 
expected redshift distributions 
 in Figure~\ref{fig:zhist}.   
The mean of all three is $z=3.9$; the dashed one is a gaussian with 
$\sigma=0.3$ and is considered 
the most likely, the dot-dashed one is a gaussian with $\sigma=0.2$ and is 
considered the narrowest possible, and the flat dotted 
one is the broadest possible (a uniform distribution from $z=3.3$ to $z=4.5$).
The presence of a bin in our redshift histogram that contains three LBGs
is only a weak indication of 
autocorrelation of the LBGs, because Poisson fluctuations of 
any of the three expected redshift distributions generate an appearance of 
clustering at least this strong 
over 10\% of the time in our Monte Carlo simulations.\footnote{Because 
this bin centered at $z=3.71$ 
is below the mean redshift, 
the result is more significant for the gaussian expected redshift 
distributions, but this 
could easily be weakened if the true mean of the expected redshift 
distribution is $z=3.8$. 
We will revisit this issue when 
the expected redshift distribution is better determined by observations 
in additional fields.}
A Kolmogorov-Smirnov 
test of the unbinned  
cumulative redshift distribution versus the three expected distributions  
fails to find any inconsistencies at the 
95\% confidence level.
A similar analysis of the 
reported redshifts of S99 
indicates that the redshift histograms of all 7 of their fields are consistent 
with Poisson fluctuations of their expected redshift distribution.

We are interested primarily in the clustering environment of the 
DLAs.  This allows us to select the bins known a priori to contain 
a DLA and to ask if a statistically significant overdensity (or deficit) of 
LBGs has been found there.  Finding zero LBGs in the bin containing the 
$z=3.86$ DLA should occur at least 40\% of the time for any of 
the expected redshift distributions.  Finding zero LBGs in the bins 
containing the  
$z=4.20$ DLA or $z=4.37$ quasar is at least as common.  Even the 
apparent triple coincidence of finding no LBGs in any of those bins occurs 
in at least 30\% of our Monte Carlo simulations, so we do not have 
good evidence for an anticorrelation of DLAs and LBGs.  Our results provide 
an upper limit on the amplitude of the 
DLA-LBG cross-correlation function, but it is not a particularly 
interesting upper limit, as even the observed strong autocorrelation function 
of $z=3$ LBGs would be difficult to detect with so few objects.  
A vast improvement in statistics will be 
required to measure the cross-correlation of 
LBGs and DLAs.


The minimum redshift uncertainty of our LBGs is 
$\Delta z=0.01$, 
so correlations on comoving scales of a few Mpc and 
smaller can only be probed via angular clustering. 
While 8 LBGs and 2 DLAs are insufficient to measure the angular 
cross-correlation precisely, we can again test whether the observed 
distribution is consistent with being random.  
The presence of 1 out of 7 LBGs from our I$_{AB}\leq 25.5$ sample within 15$''$ of the QSO line-of-sight
is inconsistent with a random distribution at the 95\% confidence 
level.\footnote{G1 was selected only because it lies within 10$''$ of the 
quasar and was therefore excluded from this analysis.}  
Since G2 is not particularly close to the QSO or DLAs in redshift this 
angular overdensity appears to be a coincidence rather than the projection 
of a three-dimensional structure.  

\section{Discussion}
\label{sect:conc}

We have not found significant evidence for autocorrelation amongst 
the 8 LBGs at $z>3.5$ in our current sample.   It is therefore not 
surprising that we find no evidence for a cross-correlation between 
LBGs and damped Lyman $\alpha$ absorbers, as the latter 
should be more difficult to detect if the DLAs are indeed less massive 
than the LBGs.  
Our results at $z\simeq 4$ are comparable to the limited studies of 
the clustering environment of damped Lyman $\alpha$ systems 
that have been performed at $z\simeq 3$.  \citet{steideletal96} found 
only one out of eleven Lyman-break galaxies in 
the field of Q0000-263 to be within $\Delta z=0.04$ of the $z=3.39$ 
DLA.  \citet{ellisonetal00} found two out of 27 LBGS in the field 
containing Q0201+1120 to be within $\Delta z=0.02$ of the $z=3.39$ 
DLA.  This failure to detect clustering of LBGs with 
DLAs at $z\simeq3$ stands in contrast to the detection of strong 
autocorrelation of LBGs at $z\simeq 3$ \citep{adelbergeretal98}, although 
the latter was determined from a much larger sample of galaxies.

In order to obtain better statistics in the future, 
we have begun a systematic survey for Lyman-break galaxies associated 
with damped Lyman $\alpha$ absorbers at $z \simeq 3$.  The 
increased sky density of LBGs bright enough for spectroscopic identification
 at $z \simeq 3$ and reduced sky noise near 
5000 \AA ~should allow us to obtain significantly improved statistics.  
The coincidence of LBG overdensity and damped Lyman $\alpha$ 
redshifts in a number of fields would argue  in favor of 
strong biasing and hence large mass of the damped Lyman $\alpha$ 
absorbers, 
which is counter to the predictions of 
the $\Lambda$CDM model.

We see no positive indication of clustering of LBGs with 
the quasar BR 0951$-$04, but this quasar is at too high a 
redshift for us to be sensitive to 
an LBG-quasar cross-correlation except in the unlikely case that
our expected redshift distribution is uniform.   
Observing fields with a quasar at lower redshift will allow us 
to measure the cross-correlation of 
quasars and LBGs, which will test the 
belief that quasars are highly 
biased markers of protoclusters \citep{djorgovski97}.
With better statistics, 
this observational program will 
measure the bias of DLAs, quasars, and 
Lyman-break galaxies, revealing much about the typical
mass and environment of these high-redshift objects.

\acknowledgments

We wish to thank Kurt Adelberger, Chuck Steidel, and Dan Stern for 
invaluable conversations 
and Jeff Newman, Andy Bunker, Max Pettini, and the referee 
for their comments which improved 
this paper.  
We thank the Keck Observatory staff for their patient and skillful  
assistance with the observations.   
We used the BOGUS package
written by Dan Stern, Andy Bunker, and Adam Stanford 
and cosmic ray removal routines written by Mark Dickinson
to reduce our multislit data.  
AMW was partially supported by NSF grant AST 0071257, and 
JXP acknowledges support from a Carnegie postdoctoral fellowship.



\begin{table}[h!]
\caption{Observing log for imaging and spectroscopy in the field of BR 0951$-$04.}  
\label{tab:obs}
\begin{tabular}{l|l|l|l|l|l}
\hline
Date& Observing Mode&Exposure Time&Seeing&$1 \sigma$ sky
&$3 \sigma$ source detection\\
	&	&	(s)&	($''$)&	AB$^a$ ~mag/sq. $''$ 
& AB mag/FWHM$^2$\\
\hline
1999 January 13	& B imaging	&4500	&0.85	&29.7	&28.7\\	
		& R imaging	&1620	&0.75	&29.0	&28.1\\
		& I imaging	&1560	&0.63	&28.1  	&27.4\\
1999 March 22 	& longslit 	&7200	&1.2	&	&\\
1999 December 6 &multislit-left	&7200 	&1.0	&	&\\
		&multislit-right&3600	&1.3	&	&\\
2000 March 9   	&multislit-left	&3600	&1.1	&	&\\
		&multislit-right&7200	&1.2	&	&\\
\hline
\end{tabular}
\end{table}
\noindent
$^a$ An AB magnitude of zero corresponds to $3.64\times10^{-23}$
 W m$^{-2}$ 
Hz$^{-1}$.  AB$_{95}$ magnitudes \citep{fukugitaetal96} 
can be converted to the Johnson-Cousins system using 
$B_{JC} = B_{AB} + 0.15$, $R_{JC} = R_{AB} - 0.18$, 
and $I_{JC} = I_{AB} - 0.43$.

\newpage

\begin{table}[t!]
\caption{Spectroscopic results for candidate $z>3.5$ galaxies.
}  
\label{tab:q0951}
\begin{tabular}{l|r|r|r|r|r|r}
\hline
Category:&	I
&II
&III
&IV
&V
& 	all\\
\hline
$z>3.5$&	3&	1&	3&	0& 	1&	8\\
$z<3.5$&       	3&      6&    	3&   	1&	0&	13\\
uncertain&     	8&      1&     	4&   	1&	0&	14\\
not observed&   5&      0&      5&   	0&	0&	10\\
\hline
Total&         	19&     8&     15& 	2&	1&	45\\
\hline
\end{tabular}
\end{table}
\noindent
I:  meets two-color criteria, I$_{AB} \leq$25\\
II:  selected based on photometric redshift, I$_{AB} \leq$25\\
III:  meets two-color criteria, $25<$I$_{AB}\leq 25.5$\\
IV:  selected based on photometric redshift, $25<$I$_{AB}\leq 25.5$\\
V:  within 10$''$ of quasar, 25.5$<$I$_{AB}\leq$26.0\\

\begin{table}[h!]
\caption{Spectroscopically confirmed $z>3.5$ galaxies.
}  
\label{tab:gals}
\begin{tabular}{l|r|r|r|r|r|r|r}
\hline
Galaxy:&	z& $\sigma_z$$^a$
&I$_{AB}$
& (B$-$R)$_{AB}$
& (R$-$I)$_{AB}$
 & z$_{prob}$$^b$
& z$_{phot}$$^c$
\\
\hline
G1& 3.69& -0.05& 25.7&      2.2&     0.0&     0.74&  3.78\\
G2& 3.81& 0.01&  24.6&      2.8&     0.0&     1.00&  3.82\\
G3& 3.72& +0.03& 24.2&      2.8&     0.0&     1.00&  3.85\\
G4& 3.57& 0.01&  24.8&1.6$^d$
					&-0.2&0.82& 3.62\\
G5& 3.75& +0.05& 25.3&      2.6&     -0.1&     0.91&  3.76\\
G6& 4.41& 0.01&  24.9&      3.2&     0.0&    0.99&  3.93\\
G7& 3.72& 0.01&  25.2&      2.3&     -0.2&     1.00&  3.71\\
G8& 4.14& +0.09& 25.5&      3.6&     0.2&     0.90&  4.08\\
\hline
\end{tabular}
\end{table}
\noindent 
$^a$ A plus or minus sign indicates that a specific 
other redshift, z $+ \sigma_z$  is also a good fit.  Otherwise, the 
errors are approximately symmetrical and roughly correspond to 
95\% confidence.\\
$^b$ Defined as the integrated probability that the redshift 
is in the range $3.5<z<4.5$.  The method corresponds to assuming a uniform 
prior as a function of redshift but then optimizing the likelihood at 
each redshift over a set of possible template spectra.\\
$^c$ The redshift which maximizes the photometric redshift 
likelihood function.\\
$^d$ G4 falls outside of the color selection region and was 
included on the basis of its photometric redshift.  The very blue $R-I$ 
color makes it fall directly on the evolutionary track of starburst 
galaxies at $z\simeq3.6$ 
which are at too low redshift to drop out 
of the $B$ filter as fully as the color criteria require.\\

\newpage




\end{document}